\documentclass[a4paper,12pt]{article}

\usepackage[table,dvipsnames,svgnames,x11names,hyperref]{xcolor}
\usepackage[utf8]{inputenc}

\usepackage{lmodern}
\usepackage[T1]{fontenc}

\usepackage[top=1in, bottom=1.25in, left=1.25in, right=1.25in]{geometry}
\usepackage{amsmath,amssymb,amsthm}

\usepackage{tikz}
\usepackage{array}
\usepackage{enumerate}
\usepackage{graphicx}
\usepackage{float}
\usepackage{caption}
\usepackage{subcaption}
\usepackage[colorlinks=true, allcolors=MidnightBlue]{hyperref}
\usepackage{rotating}
\usepackage{tikz-cd}
\usepackage{tikzpeople}

\newtheorem{thm}{Theorem}

\newtheorem{lem}[thm]{Lemma}
\newtheorem{prop}[thm]{Proposition}

\theoremstyle{definition}

\theoremstyle{remark}

\newtheorem*{example}{Example}

\newtheorem*{remarks}{Remarks}

\newcommand{\overbar}[1]{\mkern 1.5mu\overline{\mkern-1.5mu#1\mkern-1.5mu}\mkern 1.5mu}

\author{Sel\c{c}uk Kayacan}

\title{Presentations of Racks}
\date{}

\begin{document}

\maketitle

\small

\begin{center}
  Bahçeşehir University, Faculty of Engineering\\ and Natural Sciences,
  Istanbul, Turkey\\
  {\it e-mail:} \href{mailto:selcuk.kayacan@bau.edu.tr}{selcuk.kayacan@bau.edu.tr}
\end{center}

\begin{abstract}
  Presentations of racks is studied and a cryptographic protocol defined on racks is proposed.
  


\end{abstract}


\bigskip

\noindent
\textbf{Group presentations.} Let $A$ be a set of symbols and $\overbar{A}$ be the set consisting of the elements $\bar{a}$ for all $a\in A$. We define the set $A^*$ as the set of all words (finite strings) that are formed by the symbols in the symmetric set $A\cup\overbar{A}$. Let $1$ be the empty word and consider the equivalence relation $\sim$ on $A^*$ that is generated by the set of all relations $a\bar{a}\sim 1$, where $a\in A$. The \emph{free group} $\mathrm{F}(A)$ on the set $A$ is the group whose underlying set is the set of equivalence classes of $\sim$ on $A^*$ and whose group operation is the concatenation of those equivalence classes as strings. Clearly, the inverse of the word $w=ab$ in $\mathrm{F}(A)$ is the word $w^{-1}=\bar{b}\bar{a}$. The free group $\mathrm{F}(A)$ has the universal property that for any map $f$ from $A$ to a group $G$, there exists a unique group homomorphism making the following diagram commute:
\begin{center}
    \begin{tikzcd}
      A \arrow[rd, "f"'] \arrow[r, "\iota"] & \mathrm{F}(A) \arrow[d, dashrightarrow] \\
      & G
    \end{tikzcd}
  \end{center}

  Let $G$ be a group generated by the set $A$. By extending the map $f\colon A\to G$ in the obvious way to $\mathrm{F}(A)$ we see that the group $G$ is isomorphic to the quotient $\mathrm{F}(A)/N$, where $N$ is the set of all words in $\mathrm{F}(A)$ whose image is the identity element of $G$. Let $R$ be a set of relations $u_i = v_i$ in $\mathrm{F}(A)$, where $i$ runs in an index set $I$. We say $G$ has a \emph{presentation} $G = \langle A \mid R \rangle$ if $N$ is the smallest normal subgroup of $\mathrm{F}(A)$ containing the elements $u_iv_i^{-1}$, $i\in I$. The group $G$ is said to be \emph{finitely presented} if both $A$ and $R$ are finite sets.

\noindent
\textbf{Rack presentations.} A \emph{rack} is a set $X$ with a binary operation $\triangleright$ having the following two properties:
\begin{enumerate}
\item[(A1)] For all $a,b,c \in X$ we have $a\triangleright (b\triangleright c) = (a\triangleright b)\triangleright (a\triangleright c)$ (left-distributive).
\item[(A2)] For all $a,c\in X$ there exists a unique $b\in X$ such that $a\triangleright b = c$ (bijectivity).
\end{enumerate}
Rack-like structures appeared in literature in different times under various names. They are especially important in knot theory as they are useful to determine knot invariants.

We say $(X,\triangleright)$ is a \emph{quandle} if, further, the following axiom holds:
\begin{enumerate}
\item[(A3)] For all $a\in X$ we have $a\triangleright a = a$.
\end{enumerate}
Let $X$ be a subset of a group $G$ which is closed under the conjugation operation $a\triangleright b = aba^{-1}$. For $(X,\triangleright)$ all three axioms are satisfied; hence, it is a quandle. We call those type of quandles \emph{conjugation racks}. In particular, a group $G$ with the conjugation operation is a conjugation rack $(G,\triangleright)$ which we may call a \emph{group rack}. The map $\mathrm{Con}$ taking a group into the corresponding group rack is a functor from the category of groups to the category of racks. Instead of $(G,\triangleright)$ or $\mathrm{Con}(G)$ we may simply write $G$ when we want to refer to the corresponding group rack as the meaning should be clear from the context.

For a given set $A$, the \emph{free rack} $\mathrm{FR}(A)$ on $A$ is defined as $\mathrm{F}(A)\times A$ with the rack operation $$ (u,a)\triangleright (v,b) = (uau^{-1}v,b). $$
The \emph{free rack} $\mathrm{FR}(A)$ has the universal property that for any map $f$ from $A$ to a rack $X$, there exists a unique rack homomorphism making the following diagram commute:
\begin{center}
    \begin{tikzcd}
      A \arrow[rd, "f"'] \arrow[r, "\iota"] & \mathrm{FR}(A) \arrow[d, dashrightarrow] \\
      & X
    \end{tikzcd}
  \end{center}

  Let $X$ be a rack generated by the set $A$ and $R$ be a set of relations $(u_i,a) = (v_i,b)$ in $\mathrm{FR}(A)$, where $i$ runs in an index set $I$. Clearly, the equivalence classes of $\mathrm{FR}(A)$ form a rack with the induced rack operation. We say $X$ has a \emph{presentation} $X = \langle A \mid R \rangle_{\mathrm{rk}}$ if $X$ is isomorphic to the rack of those equivalence classes. The rack $X$ is said to be \emph{finitely presented} if both $A$ and $R$ are finite sets. A particular example is the \emph{free quandle} $\mathrm{FQ}(A)$, where the equivalence relation $R_q$ on $\mathrm{FR}(A)$ is generated by the set of all relations $(w,a) = (wa,a)$, where $a\in A$ and $w\in \mathrm{F}(A)$. Observe that the map $(w,a)\mapsto waw^{-1}\colon \mathrm{FQ}(A)\to \mathrm{F}(A)$, where we regard $\mathrm{F}(A)$ as a conjugation rack, is an injective rack homomorphism. Thus, the free quandle $\mathrm{FQ}(A)$ is isomorphic to the conjugation rack which is the union of conjugacy classes of the elements of $A$ in $\mathrm{F}(A)$.

Given a rack $X$ its \emph{enveloping group} (or its \emph{associated group}) is defined as $$ \mathrm{Env}(X) = \langle X \mid ab\bar{a} = a\triangleright b,\; a,b\in X \rangle. $$
The map $\mathrm{Env}$ is a functor from the category of racks to the category of groups. Actually, $\mathrm{Env}$ is left-adjoint to the functor $\mathrm{Con}$.

\begin{thm}[see {\cite[Lemma~4.3]{FR92}}]\label{thm:env}
  Let $X=\langle A\mid R \rangle_{\mathrm{rk}}$ be a finitely presented rack. Then $\langle A\mid R\rangle$ is a finite presentation for the group $\mathrm{Env}(X)$.
\end{thm}

\noindent
\textbf{Rack automorphisms.} Let $X$ be a rack. The two defining axioms (A1) and (A2) implies that, for any $a\in X$, the left-multiplication map $\phi_a\colon b\mapsto a\triangleright b$ is an automorphism of $X$. Let $\Phi$ be the map taking $a\in X$ to the automorphism $\phi_a$. Then the set $\Phi(X)$ together with the operation $\phi_a\triangleright \phi_b = \phi_a\phi_b\phi_a^{-1}$ is a conjugation rack. Actually, the map $\Phi\colon X\to \Phi(X)$ is a rack homomorphism as $\Phi(a\triangleright b) = \Phi(a)\triangleright \Phi(b)$. The \emph{inner automorphism group} $\mathrm{Inn}(X)$ is defined as the normal subgroup of the symmetric group $\mathrm{Sym}(X)$ generated by the automorphisms $\phi_a$, $a\in X$; i.e., $$ \mathrm{Inn}(X) = \langle \Phi(X) \rangle. $$

\begin{prop}[see {\cite[Lemma~1.7]{AG03}}]\label{prop:inn}
  Let $X\subseteq G$ be a conjugation rack. Then the map $\Phi\colon X\to \Phi(X)$ extends to a surjective group homomorphism
  $$\Phi\colon G\to \mathsf{Inn}(X),$$
  whose kernel is the center $\mathrm{Z}(G)$ of $G$.
\end{prop}

Given a rack $X$ and an element $w\in \mathrm{F}(X)$, we denote by $\phi_w$ the composition of the automorphisms $\phi_{a_i}^{\pm 1}$, $a_i\in X$, following the string expansion of $w$. For example, if $w=a_1\bar{a}_2$ then $\phi_w = \phi_{a_1}\phi_{a_2}^{-1}$. The group $\langle X \mid R \rangle$ in the statement of the below Theorem is called the \emph{operator group} of $X$ in \cite{FR92}.

\begin{thm}
  Let $X$ be a rack. Then the inner automorphism group $\mathrm{Inn}(X)$ is isomorphic to the group $$ \langle X \mid R \rangle, $$
  where $R$ is the set of all relations $w=1$ such that $\phi_w = \mathrm{id}_X$.
\end{thm}

\noindent
\textbf{Rack extensions.} We say $X$ is a \emph{connected} rack if the action of $\mathrm{Inn}(X)$ on $X$ is transitive.

\begin{lem}[see {\cite[Lemma~1.21]{AG03}}]\label{lem:fiber}
    Let $X$ and $Y$ be two racks and let $f\colon X\to Y$ be a surjective rack homomorphism. Then, for any $y_1,y_2\in Y$, the cardinalities of the fibers $f^{-1}(y_2)$ and $f^{-1}(y_1\triangleright y_2)$ are equal. In particular, every fiber of $Y$ has the same cardinality if $Y$ is connected.
\end{lem}

\begin{prop}[see {\cite[Lemma~2.1]{AG03}}]
  Let $Y$ be a rack and $F$ be a set. Let $\alpha$ be a family of functions $\alpha_{y_1y_2}\colon F\times F \to F$ indexed by the pairs of elements of $Y$. Consider the set $Y\times F$ with the operation
  $$  (y_1,i)\triangleright (y_2,j) := (y_1\triangleright y_2, \alpha_{y_1y_2}(i,j))  $$
  Then $Y\times F$ is a rack if for every $y_1,y_2,y_3\in Y$ and $i,j,k\in F$
  \begin{itemize}
  \item $\alpha_{y_1y_2}(i,\cdot)\colon F\to F$ is a bijection, and the equality
  \item $\alpha_{y_1\phi_{y_2}(y_3)}(i,\alpha_{y_2y_3}(j,k)) = \alpha_{\phi_{y_1}(y_2)\phi_{y_1}(y_3)}(\alpha_{y_1y_2}(i,j),\alpha_{y_1y_3}(i,k))$ holds.
  \end{itemize}
  Furthermore, the rack $Y\times F$ is a quandle if $Y$ is a quandle and, additionally, for every $y \in Y$ and $i\in F$ the equality
  \begin{itemize}
  \item $\alpha_{yy}(i,i) = i$ holds.
  \end{itemize}
\end{prop}

We write $Y\times_{\alpha} F$ if $\alpha$ satisfies the first two conditions specified in the above Proposition and call $Y\times_{\alpha} F$ an \emph{extension} of $Y$.

\begin{thm}[see {\cite[Corollary~2.5]{AG03}}]
 Let $f\colon X\to Y$ be a surjective rack homomorphism such that all fibers $f^{-1}(y)$ have the same cardinality. Then $X$ is an extension $Y\times_{\alpha} F$.
\end{thm}

\noindent
\textbf{Protocol Description.} Alice and Bob wants to form a shared secret without prior knowledge. This shared secret can be used to derive a symmetric secret key so that their later communication can be encrypted with the symmetric secret key. To form the shared secret we propose a protocol defined on a rack $X$. The public parameters are $X$ and some set of elements $x_1,x_2,\dots,x_t\in X$ generating a large subset of $X$. Alice chooses a secret element $a\in X$ and computes $\phi_a(x_1),\dots,\phi_a(x_t)$. We assume the existence of a Trusted Third Party (TTP) that binds the identity of Alice with the public knowledge $\phi_a(x_1),\dots,\phi_a(x_t)$ through a certificate and Bob already owns Alice's certificate. To derive a shared secret Alice picks an element $x\in X$ at random, computes $\phi_a(x)$ and sends $x$ together with $\phi_a(x)$ to Bob. Upon receiving the message Bob computes $\phi_b(x)$ and sends it to Alice. Here, the element $b\in X$ must be expressible in terms of $x_i$, $1\leq i\leq t$. It may be an ephemeral value that is produced at random upon receiving a message from Alice; or, it may be a static value that is used to form Bob's certificate. At the end of the message exchange both parties can compute the shared secret $ \phi_a\phi_b(x) = \phi_{\phi_a(b)}\phi_a(x)$: Clearly, Alice can compute $\phi_a\phi_b(x)$ as she knows $a$ and received $\phi_b(x)$ from Bob. Bob can also compute the shared secret as he knows $\phi_a(x_1),\dots,\phi_a(x_t)$ through Alice's certificate which means he can compute $\phi_a(b)$.

\begin{figure}
  \begin{center}
    \begin{tikzpicture}[every node/.style={minimum width=1.5cm}]
      \node[alice] (A) {$\phi_a\phi_b(x)$};
      \node[bob, right= 4cm of A, mirrored] (B) {$\phi_{\phi_a(b)}\phi_a(x)$};
      \draw (A.35) edge[->,thick] node[above] {$x, \phi_a(x)$} (B.145);
      \draw (A.345) edge[<-,thick] node[above] {$\phi_b(x)$} (B.195);
    \end{tikzpicture}  
  \end{center}
\end{figure}

\begin{remarks}
  \mbox{}
  \begin{itemize}
  \item For this protocol to work there should be an efficient algorithmic procedure whose output is a standard form of the input value $x\in X$ so that both parties would end up with the same bit sequence after the message exchange. 
  \item For the security of the protocol, given $x,x_1,\dots,x_t\in X$ and their images under $\phi_a$, it should be computationally infeasible to find $a\in X$.
  \end{itemize}
\end{remarks}

\noindent
\textbf{Parameter selection.} The obvious way to introduce the rack $X$ as a public parameter in the above protocol is to describe $X$ via one of its presentations. Let $A$ be a countable set and consider the free group $\mathrm{F}(A)$. One can easily describe a procedure to specify a subset $\mathrm{F}(A)^+$ of $\mathrm{F}(A)$ such that a non-empty word $w$ lies in $\mathrm{F}(A)^+$ if and only if $w^{-1}$ does not lie in $\mathrm{F}(A)^+$. Notice that $\mathrm{FR}(\mathrm{F}(A)^+)$ is isomorphic to $\mathrm{Con}(\mathrm{F}(A))$. If we define $X$ as a quotient of $\mathrm{FR}(\mathrm{F}(A)^+)$, in the protocol description the element $b$ that is selected by Bob can be taken from the group generated by the elements $x_i$, $1\leq i\leq t$, where $x_i$ are regarded as the elements of $\mathrm{F}(A)$.

\begin{example}
  The Thompson's group can be defined as $$ F = \langle a_0,a_1,a_2,\dots \mid a_ia_{j+1}a_i^{-1} = a_j\; (i<j)\rangle.  $$
  Then it is straightforward to define $\mathrm{Con}(F)$ as a quotient of the free quandle $\mathrm{FQ}(\mathrm{F}(A)^+)$ for a suitable choice of $\mathrm{F}(A)^+$.
\end{example}

It is also possible to start with a finite group $G = \langle A\mid R\rangle$ and take $X$ as a finite extension of $\mathrm{Con}(G)$. To be more precise, we suggest the following: The relations in $R$ either specifies the orders of the elements (for example, $a_1^{n_1} = 1$) or describes the action of $\mathrm{Inn}(X)$ on $X$ (for example, $\phi_{a_1}(a_2) = a_3$). In such a case elements of $X$ can be easily expressed in a standard form. However, depending on the presentation of $G$ an adversary may successfully compute $\phi_a$ with the publicly available data by  exploiting the arithmetic conditions.


\end{document}